\documentclass[12pt,a4paper]{article}

\usepackage[utf8]{inputenc}
\usepackage[T1]{fontenc}
\usepackage{lmodern}
\usepackage[english]{babel}
\usepackage{csquotes}
\usepackage{microtype}
\usepackage{chngcntr}

\usepackage[margin=2cm]{geometry}
\usepackage{setspace}
\usepackage{amsmath}
\usepackage{amssymb}
\usepackage{amsthm}
\newcommand{\see}{see, e.g. }
\usepackage{booktabs}
\usepackage{graphicx}
\usepackage{float}
\usepackage{caption}
\usepackage{subcaption}
\usepackage{multirow}
\usepackage[backend=bibtex,style=authoryear,maxnames=999,maxcitenames=3]{biblatex}
\renewbibmacro{in:}{}

\usepackage[colorlinks=true,
linkcolor=blue,
citecolor=blue,
urlcolor=blue]{hyperref}

\usepackage{enumitem}  % Better lists
\usepackage{siunitx}   % Units formatting
\usepackage{xcolor}    % Colors
\usepackage{todonotes} % Comments and todos
\usepackage{pdflscape} % Provides the landscape environment

\begin{document}
\nocite{*}

\def\spacingset#1{\renewcommand{\baselinestretch}%
{#1}\small\normalsize} \spacingset{1}

%%%%%%%%%%%%%%%%%%%%%%%%%%%%%%%%%%%%%%%%%%%%%%%%%%%%%%%%%%%%%%%%%%%%%%%%%%%%%%

{
   \title{\bf A difference-in-differences estimator by covariate balancing propensity score}
  \author{Junjie Li%\thanks{The authors gratefully acknowledge \textit{please remember to list all relevant funding sources in the unblinded version}}\hspace{.2cm}
  \\
    Department of Economics, Hitotsubashi University \\
    %\href{mailto:ed235009@g.hit-u.ac.jp}{	ed235009@g.hit-u.ac.jp}\\
    and \\
    Yukitoshi Matsushita \\
    Department of Economics, Hitotsubashi University\\
    %\href{mailto:matsushita.y@r.hit-u.ac.jp}{	matsushita.y@r.hit-u.ac.jp}\\
    }

  \maketitle
}

\bigskip
\begin{abstract}
This article develops a covariate balancing approach for the estimation of treatment effects on the treated (ATT) in a difference-in-differences (DID) research design when panel data are available. We show that the proposed covariate balancing propensity score (CBPS) DID estimator possesses several desirable properties: (i) local efficiency, (ii) double robustness in terms of consistency, (iii) double robustness in terms of inference, and (iv) faster convergence to the ATT compared to the augmented inverse probability weighting (AIPW) DID estimators when both working models are locally misspecified. These latter two characteristics set the CBPS DID estimator apart from the AIPW DID estimator theoretically. Simulation studies and an empirical study demonstrate the desirable finite sample performance of the proposed estimator. 
\end{abstract}

\noindent%
\vfill

\newpage
\spacingset{1.8} % DON'T change the spacing!
\section{Introduction}
\hspace{1em} Difference-in-differences (DID) is a widely employed research design in evaluating the causal effects of policy interventions using observational data. In its canonical form, the DID approach necessitates two groups and two periods, stipulating that no entity is exposed to the treatment in the initial period, while a subset remains untreated in the subsequent period. A crucial underpinning of the DID design is the so-called (unconditional) parallel trends assumption (PTA), which posits that, in the absence of the treatment, the average outcomes for both the treatment and comparison groups would have evolved along parallel paths over time. While the PTA is inherently untestable, its validity has been questioned, particularly in scenarios where pre-treatment characteristics, differing between the treatment and comparison groups, are correlated with the outcome's evolution. In such instances, the canonical DID setup becomes implausible, prompting researchers to incorporate pre-treatment covariates into the DID framework. This modification ensures the satisfaction of the PTA conditionally on these covariates (conditional PTA). 
 
Under this conditional PTA, three predominant estimation procedures have emerged: the outcome regression (OR), the inverse probability weighting (IPW), and the augmented IPW (AIPW), the latter offering double robustness in terms of consistency, see \textcite{sant2020doubly} for a comprehensive review. However, these methods confront the challenge of potential misspecification of the outcome regression and/or propensity score models, leading to incorrect inferences. While doubly robust estimators offer improved consistency by requiring only one of the two working models to be correctly specified, an unavoidable situation where both models are misspecified still remains. For estimation of the average treatment effect (ATE), \textcite{kang2007demystifying} highlight this limitation, demonstrating that the advantages of AIPW estimators can substantially diminish when both the outcome regression and propensity score models are slightly misspecified. To address this issue, \textcite{imai2014covariate} proposed the Covariate Balancing Propensity Score (CBPS) methodology, illustrating that the CBPS estimator can significantly ameliorate the finite sample performance of doubly robust estimators. Further theoretical expositions of the CBPS ATE estimator were provided by \textcite{fan2022optimal}. Nevertheless, their investigation focuses on ATE estimation. 

In this paper, we apply the CBPS methodology to the ATT estimator within the framework of DID design. 
In particular, we rigorously investigate the robustness and efficiency properties of the CBPS DID estimator. 
Our contributions to the DID literature are manifold: Firstly, this article briefly reviews a range of existing ATT estimators within the DID framework and then propose a CBPS-based DID estimator when panel data are available. Secondly, we demonstrate that our proposed estimator possesses the qualities of double robustness and local efficiency. A notable distinction of our estimator is its double robustness not only in terms of consistency but also in terms of inference. This characteristic guarantees that the  asymptotic linear representation of the CBPS DID estimator remains invariant even when one of the working models is misspecified. As a consequence, it allows us to estimate the asymptotic variance based on the influence function. This feature sets our estimator apart from existing doubly robust AIPW DID estimators because the asymptotic linear representation of the AIPW DID estimator is not invariant when one of the working models is misspecified. The third contribution of our work lies in establishing that our estimator can achieve a faster convergence rate relative to the AIPW DID estimator, under the scenario where both the propensity score and outcome regression models are locally misspecified. This situation has been seldom addressed in existing DID literature. Lastly, the fourth contribution is the practicality of our estimator. It is straightforward to implement. This simplicity in application enhances its utility in empirical research.

\textbf{Organization of this paper:} The subsequent section of this paper delineates the settings and assumptions used throughout the paper and briefly overviews some existing DID estimators. In Section 3, we introduce the CBPS method and propose a CBPS-based DID estimator, and derive its theoretical properties. We evaluate the finite sample performance of our proposed CBPS DID estimator with Monte Carlo simulation in Section 4, and provide an empirical example in Section 5. Conclusions are summarized in Section 6. All mathematical proofs supporting our arguments and findings are comprehensively compiled in the Appendix.

\section{Difference-in-differences}
\subsection{Setup}
\hspace{1em} We introduce the notation and framework that will be employed throughout this article. Our analysis is based on a two-period, two-group structure (treatment and comparison groups). Let $Y_{it}$ represent the outcome of interest for unit $i$ at time $t$. We suppose that researchers have access to outcome data both in a pre-treatment period $t=0$ and in a post-treatment period $t=1$. Define $D_{it}$ as an indicator variable, where $D_{it}=1$ if unit $i$ receives treatment on time $t$, and $D_{it}=0$ otherwise. We note that $D_{i0}=0$ for all units $i$, which simplifies the treatment indicator to $D_i=D_{i1}$. The observed outcome $Y_{it}$ can also be rewritten as $Y_{it}=D_{i}Y_{it}(1)+(1-D_i)Y_{it}(0)$, where $Y_{it}(0)$ denotes the potential outcome of unit $i$ at time $t$ if one does not receive treatment and $Y_{it}(1)$ represents the potential outcome if the same one receives treatment, but $Y_{it}(0)$ and $Y_{it}(1)$ cannot be observed simultaneously for any unit. In the remainder of this paper, we assume the availability of panel data on $\{Y_{i0}, Y_{i1}, D_i, X_i\}_{i=1}^n$, where $X_i \in \mathbb{R}^{d}$ is a vector of pre-treatment covariates and the first element of $X_i$ is a constant.\\ 

The parameter of interest, the average treatment effect on the treated (ATT), is defined as:
\begin{equation*}
\tau=E\left[Y_{i1}(1)-Y_{i1}(0)|D_i=1\right]\tag{2.1}. \label{ATT def}
\end{equation*}
Since $Y_{i1}(1)=Y_{i1}$ given $D_i=1$, ATT can be rewritten as: 
\begin{equation}
\tau=E\left[Y_{i1}|D_i=1\right]-E\left[Y_{i1}(0)|D_i=1\right]. \tag{2.2} \label{ATT change}
\end{equation}
According to the representation (\ref{ATT change}) above, we can show that the first term ($E\left[Y_{i1}|D_i=1\right]=\frac{E[D_i Y_{i1}]}{E[D_i]}$) can be estimated directly from the observed data. The main challenge in identifying the ATT lies in computing the second term ($E\left[Y_{i1}(0)|D_i=1\right]$) from the observed data since $Y_{i1}(0)$ is missing for treated subjects with $D_i=1$.
In order to identify the ATT (or $E\left[Y_{i1}(0)|D_i=1\right]$), the following assumptions are necessary.\\

\noindent\textbf{Assumption 1.} Assume that the data $\{Y_{i0},Y_{i1},D_i,X_i\}_{i=1}^n$ are independent and identically distributed ($iid$).\\

\noindent\textbf{Assumption 2.} $E[Y_{i1}(0)-Y_{i0}(0)|X_i,D_i=1]=E[Y_{i1}(0)-Y_{i0}(0)|X_i,D_i=0]$ almost surely.\\

\noindent\textbf{Assumption 3.} For some $\varepsilon>0$, $\Pr(D_i=1)>\varepsilon$ and $\Pr(D_i=1|X_i)\leq 1-\varepsilon$ almost surely.\\

Assumption 2, which we term the conditional PTA, posits that the average conditional outcomes for the treatment and comparison groups would have followed parallel paths in the absence of the treatment. Assumption 3 is an overlap condition, which states that at least a small fraction of the population is exposed to the treatment, and for every specific value of covariates $X_i$, at least a small portion is not treated. These three assumptions are commonly utilized in semiparametric DID methods, \see \parencites{heckman1997matching}{abadie2005semiparametric}{sant2020doubly}. Next, we briefly provide an overview of the existing approaches to identify and estimate the ATT.

\subsection{Existing DID estimators}
\hspace{1em} There are two approaches to identify the ATT: the OR approach (\textcite{heckman1997matching}) and the IPW approach (\textcite{abadie2005semiparametric}).\\ 

(i) The OR approach: under Assumptions 1-3, the ATT is identified as: 
 \begin{equation*}
     \tau=E\left[\Delta Y_i|D_i=1\right]-E\left[E\left[\Delta Y_i|X_i,D_i=0\right]|D_i=1\right]=E\left[\Delta Y_i|D_i=1\right]-E\left[m_{\Delta}(X_i)|D_i=1\right] :=\tau^{OR}, \tag{2.3} \label{ATT OR}
 \end{equation*}
where $\Delta Y_i=Y_{i1}-Y_{i0}$ and $m_{\Delta}(X_i)=E\left[\Delta Y_i|X_i,D_i=0\right]$. Based on the identification result (\ref{ATT OR}), the OR approach requires researchers to model the conditional expectation of outcome evolution $E\left[\Delta Y_{i}|X_i,D_i=0\right]$ at the first step. Researchers typically adopt a linear parametric model $X_i^{\prime} \gamma$ (outcome regression model) to specify the true conditional expectation of outcome evolution $E\left[\Delta Y_{i}|X_i,D_i=0\right]$. Consequently, the OR DID estimator is represented as follows:\\
\begin{equation}
\hat{\tau}^{OR}=\frac{1}{n_{treat}}\sum_{i \in treat} \Delta Y_i-\frac{1}{n_{treat}}\sum_{i \in treat}  X_i^{\prime}\hat{\gamma}^{OLS}, \tag{2.4} \label{OR est}
\end{equation}
where $\hat{\gamma}^{OLS}$ is the OLS estimator of the regression of $Y_i$ on $X_i$ by the comparison groups ($D_i=0$) and $n_{treat}$ denotes the treatment group size.\\

(ii) The IPW approach: under Assumptions 1-3, the ATT is alternatively identified as: 
\begin{equation*}
\tau=E\left[\frac{D_i-\pi(X_i)}{E[D_i](1-\pi(X_i))}\Delta Y_i\right] :=\tau^{IPW}, \tag{2.5} \label{ATT IPW}
\end{equation*}
where $\pi(X_i)=\Pr(D_i=1|X_i)$ is the propensity score. Based on the identification result (\ref{ATT IPW}), the IPW approach requires researchers to estimate the propensity score at the first step. Researchers typically use a parametric model (e.g. $\pi(X_i^{\prime}\beta)=\frac{\exp(X_i^{\prime} \beta)}{1+\exp(X_i^{\prime}\beta)}$) to specify the propensity score $\pi(X_i)$ and estimate parameters by the maximum likelihood method. Hence, the IPW DID estimator is expressed as:
\begin{equation}
{\hat{\tau}}^{IPW}=\frac{1}{n}\sum_{i=1}^n \frac{D_i-\pi(X_i^{\prime}\hat{\beta}^{ML})}{\bar{D}(1-\pi(X_i^{\prime}\hat{\beta}^{ML}))}\Delta Y_i, \tag{2.6} \label{IPW est}
\end{equation}
where $\hat{\beta}_{ML}$ is the maximum likelihood estimator and $\bar{D}=\frac{1}{n}\sum_{i=1}^{n}D_i$.\\

(iii) AIPW approach: Consistency of the OR and IPW estimators depends on the correct specification of the outcome regression model and the propensity score model, respectively. 
To achieve consistency in scenarios where one of two working models are misspecified, \textcite{sant2020doubly} proposed the following AIPW DID estimator:
\begin{equation}
\hat{\tau}^{AIPW}=\frac{1}{n}\sum_{i=1}^n \frac{D_i-\pi(X_i^{\prime}\hat{\beta}^{ML})}{\bar{D}(1-\pi(X_i^{\prime}\hat{\beta}^{ML}))}(\Delta Y_i- X_i^{\prime}\hat{\gamma}^{AIPW}),\tag{2.7} \label{AIPW est}
\end{equation}
with $\hat{\gamma}^{AIPW}=\hat{\gamma}^{OLS}$, which follows from the alternative identification of ATT:
\begin{equation}
    \tau=E\left[\frac{D_i-\pi(X_i)}{E[D_i](1-\pi(X_i))}(\Delta Y_i-m_{\Delta}(X_i))\right] :=\tau^{AIPW}.\tag{2.8} \label{ATT AIPW}
\end{equation}

Observing the form of (\ref{AIPW est}), it is apparent that AIPW procedure combines both OR and IPW methodologies. This synthesis allows the AIPW estimator to mitigate some of the inherent weaknesses associated with the OR and IPW approaches individually. Indeed, \textcite{sant2020doubly} show that the AIPW DID estimator is both locally efficient and doubly robust in terms of consistency. In the cross-sectional setting, however, \textcite{kang2007demystifying} demonstrated that the performance of the AIPW ATE estimator can be poor in scenarios where both of the working models are slightly misspecified. To address this issue, \textcite{imai2014covariate} introduced the CBPS method and demonstrated that the CBPS ATE estimator can significantly enhance performance over other existing ATE estimators, including the AIPW ATE estimator, particularly when both working models are misspecified. In the subsequent subsection, we extend the application of the CBPS method from estimating ATE to ATT within the framework of DID research. Our objective is to propose a CBPS DID estimator and to rigorously investigate its theoretical properties, focusing on their robustness and efficiency. \\

\section{CBPS DID estimator}    

\subsection{CBPS methodology}
\hspace{1em} The CBPS method introduced by \textcite{imai2014covariate} offers a distinct approach to propensity score estimation. In contrast to the IPW method, the CBPS method imposes exact finite sample balance of pre-treatment covariates across the treatment and comparison groups rather than focusing on the predictive accuracy of treatment assignment. 
The CBPS DID estimator is defined as 
 \begin{equation}
      \hat{\tau}^{CBPS}=\frac{1}{n}\sum_{i=1}^n\frac{D_i-\pi(X_i^{\prime}\hat{\beta}^{CBPS})}{\bar{D}(1-\pi(X_i^{\prime}\hat{\beta}^{CBPS}))}\Delta Y_i,\tag{3.1} \label{CBPS est}
 \end{equation}
 where $\hat{\beta}$ satisfies
 \begin{equation}
 \frac{1}{n} \sum_{i=1}^{n} \left(D_i-\frac{(1-D_i)\pi(X_i^{\prime}\hat{\beta}^{CBPS})}{1-\pi(X_i^{\prime}\hat{\beta}^{CBPS})}\right)X_i=\frac{1}{n} \sum_{i=1}^{n} \frac{D_i-\pi(X_i^{\prime}\hat{\beta}^{CBPS})}{1-\pi(X_i^{\prime}\hat{\beta}^{CBPS})} X_i=0.\tag{3.2} \label{CBPS score fun}
\end{equation}

Recall that the IPW DID estimator (\ref{IPW est}) employs the maximum likelihood method to estimate the propensity score, where $\hat{\beta}^{ML}$ satisfies $\frac{1}{n}\sum_{i=1}^n \frac{D_i-\pi(X_i^{\prime}\hat{\beta}^{ML})}{1-\pi(X_i^{\prime}\hat{\beta}^{ML})}\frac{\dot{\pi}(X_i^{\prime}\hat{\beta}^{ML})}{\pi(X_i^{\prime}\hat{\beta}^{ML})}X_i=0$ where $\dot{\pi}(v)=\frac{\partial \pi(v)}{\partial v}$. Hence the only difference of the IPW DID estimator (\ref{IPW est}) and the CBPS DID estimator (\ref{CBPS est}) is the method of propensity score estimation. On the other hand, by (\ref{CBPS score fun}), the CBPS DID estimator can also be expressed as
\begin{equation}
     \hat{\tau}^{CBPS}= \frac{1}{n}\sum_{i=1}^n\frac{D_i-\pi(X_i^{\prime}\hat{\beta}^{CBPS})}{\bar{D}(1-\pi(X_i^{\prime}\hat{\beta}^{CBPS}))}(\Delta Y_i-X_i^\prime \gamma^{CBPS}) \tag{3.3} \label{CBPS Est Prop}
\end{equation}
where $\gamma^{CBPS}$ is any value. Hence the only difference of the AIPW DID estimator (\ref{AIPW est}) and the CBPS DID estimator (\ref{CBPS est}) is the value of $\gamma$ in (\ref{AIPW est}) and (\ref{CBPS Est Prop}).
It is noteworthy that $\gamma^{CBPS}$  in (\ref{CBPS Est Prop}) can take any value and it is not estimated in the actual CBPS estimation defined as (\ref{CBPS est}). In the following subsections, we will conduct a comprehensive theoretical analysis of the CBPS DID estimator to elucidate its advantages. It is this arbitrariness of $\gamma^{CBPS}$ in (\ref{CBPS Est Prop}) that plays a key role in showing robustness to misspecification of parametric working models compared to the AIPW DID estimator.

\subsection{Local efficiency}
 \hspace{1em} In this subsection, we start with the scenario when both of the working models are correctly specified. We show that, in such a case, the CBPS DID estimator attains the semiparametric efficiency bound for the ATT under DID framework, when both propensity score model and outcome regression model are correctly specified. This property is the so-called local efficiency. \textcite{sant2020doubly} derived the semiparametric efficiency bound for ATT under a DID framework. %This bound is particularly noteworthy as it does not necessitate additional knowledge regarding the functional forms of outcome regression or propensity score model, see section 2 of Sant^{\prime}Anna and Zhao (2020) for more details.
%Before discussing how our proposed estimator achieves the semiparametric efficiency bound, we need to introduce some additional notation. 
%Let $\beta_0$ and $\gamma_0$ denote the probability limits of $\hat{\beta}$ and $\hat{\gamma}$, respectively. 
%Given that the $\gamma$ in (\ref{CBPS Est Prop}) can take any value, we can set it as $\gamma=\gamma_0$ in this section.
\\

\noindent \textbf{Theorem 1.} Under Assumptions 1-3 and Assumptions A (stated in Appendix A), if $\pi(X_i^{\prime}\beta)=\pi(X_i)$ and $X_i^{\prime}{\gamma}=m_{\Delta}(X_i)$ holds,
    \begin{equation*}
        \sqrt{n}(\hat{\tau}^{CBPS}-\tau) = \frac{1}{\sqrt{n}}\sum_{i=1}^{n}\eta_i^e+o_p(1)\tag{3.4} \label{Theorem 1}
         \xrightarrow{d} N\left(0,E[{\eta_i^e}^2]\right),
    \end{equation*}
 where 
\begin{equation*}
\eta_i^{e} =\frac{D_i-\pi(X_i)}{E[D_i]\{1-\pi(X_i)\}}(\Delta Y_i - m_{\Delta}(X_i))-\frac{D_i}{E[D_i]}\tau \tag{3.5} \label{eff inf fun}
\end{equation*}
is the efficient influence function for the ATT and $E[{\eta_i^{e}}^2]$ is the semiparametric efficiency bound.

Theorem 1 shows asymptotic normality of the CBPS DID estimator when both of the working models are correctly specified. The asymptotic variance of the CBPS DID estimator is equal to the semiparametric efficiency bound derived by \textcite{sant2020doubly}. It should be noted that the AIPW DID estimator also achieves the semiparametric efficiency bound when both of the working models are correctly specified.

\subsection{Double robustness}
\hspace{1em} In the previous subsection, we showed that the CBPS DID estimator is locally efficient. In this subsection, we shift our focus from efficiency to robustness, under the scenario that one of the two working models is misspecified. Firstly, we show that the CBPS DID estimator remains consistent with the ATT even if either the propensity score model or the outcome regression model (but not both) is misspecified. This property is referred to as double robustness in terms of consistency.
\\

\noindent\textbf{Theorem 2.} Under Assumptions 1-3 and Assumptions A, the CBPS DID estimator is doubly robust in terms of consistency, that is $\hat{\tau}^{CBPS} \xrightarrow{p} \tau$ if at least one of the following two conditions holds:\\

1. The outcome regression model is correctly specified, that is, there exists some value $\gamma_0$ such that $X_i^{\prime}{\gamma_0}=m_{\Delta}(X_i) $ a.s. \\

2. The propensity score model is correctly specified, that is, there exists some value $\beta_0$ such that  $\pi(X_i^{\prime}\beta_0)=\pi(X_i)$ a.s.\\

Consequently, the CBPS DID estimator offers more flexibility and is less demanding in terms of a researcher's ability to correctly specify nuisance parametric models, compared to the OR and the IPW approach. It should be noted that the AIPW DID estimator is also doubly robust in terms of consistency. 

While double robustness in terms of consistency is a valuable property, it may not suffice for inference. The next theorem shows that the asymptotic linear representation of the CBPS DID estimator remains invariant even when one of the working models is misspecified so that the asymptotic variance can be estimated based on the influence function. This is referred as double robustness in terms of inference. In contrast, the asymptotic linear representation of the AIPW DID estimator is not invariant when one of the working models is misspecified. A detailed proof is provided in the Appendix.\\

\noindent\textbf{Theorem 3.} Let $\beta_0^{AIPW}$, $\gamma_0^{AIPW}$, $\beta_0^{CBPS}$ and $\gamma_0^{CBPS}$ denote probability limits of $\hat{\beta}^{ML}$, $\hat{\gamma}^{AIPW}$, $\hat{\beta}^{CBPS}$ and $\hat{\gamma}^{CBPS}=\left[\sum_{i=1}^{n}\frac{(1-D_i)\Dot{\pi}(X_i^\prime \hat{\beta}^{CBPS})}{(1-\pi(X_i^{\prime}\hat{\beta}^{CBPS}))^2} X_i X_i^\prime\right]^{-1}\sum_{i=1}^{n}\frac{(1-D_i)\Dot{\pi}(X_i^\prime \hat{\beta}^{CBPS})\Delta Y_i}{(1-\pi(X_i^{\prime}\hat{\beta}^{CBPS}))^2}X_i$, respectively.
Under Assumptions 1-3 and Assumptions A, if either $\pi(X_i^{\prime}\beta_0^a)=\pi(X_i)$ a.s. or $X_i^{\prime}{\gamma_0^a}=m_{\Delta}(X_i)$ a.s. for $a=CBPS, AIPW$, the CBPS DID and AIPW DID estimators satisfy:
\begin{align*}
    &\sqrt{n}(\hat{\tau}^{CBPS}-\tau)=\frac{1}{\sqrt{n}}\sum_{i=1}^{n}\eta_i^{CBPS}+o_p(1), \tag{3.6} \label{CBPS asy rep}\\
    & \sqrt{n}(\hat{\tau}^{AIPW}-\tau)=\frac{1}{\sqrt{n}}\sum_{i=1}^{n}\eta_i^{AIPW} +O_p(1), \tag{3.7} \label{AIPW asy rep}
\end{align*}
where $\eta_i^a=\frac{D_i-\pi({X_i}^{\prime}\beta_0^a)}{E[D_i]\{1-\pi({X_i}^{\prime}\beta_0^a)\}}(\Delta Y_i -X_i^{\prime}{\gamma_0^a})-\frac{D_i}{E[D_i]}\tau$. Note that $\eta_i^a$ is equal to the efficient influence function (\ref{eff inf fun}) under the assumption that both working models are correctly specified.

Theorem 3 reveals that inference based on $\hat{\tau}^{AIPW}$ and its influence function may be misleading when one of the working models is misspecified. In contrast, inference based on $\hat{\tau}^{CBPS}$ and its influence function will remain accurate even when one of the working models is misspecified. This double robustness in terms of inference significantly enhances the appeal of the CBPS DID estimator. %Consequently, the form of asymptotic variance of the CBPS DID estimator remains invariant, regardless of whether the propensity score model or the outcome regression model is misspecified. 
%Practically, this attribute facilitates simpler and more reliable inference procedures. 
We note that although $\hat{\gamma}^{CBPS}$ does not appear in estimating $\hat{\tau}^{CBPS}$ (see (\ref{CBPS est})), it does appear in estimating the asymptotic variance. Specifically, the asymptotic variance of the CBPS DID estimator should be estimated by $\frac{1}{n}\sum_{i=1}^n\left\{\frac{D_i-\pi({X_i}^{\prime}\hat{\beta}^{CBPS})}{\bar{D}\{1-\pi({X_i}^{\prime}\hat{\beta}^{CBPS})\}}(\Delta Y_i -X_i^{\prime}{\hat{\gamma}^{CBPS}})-\frac{D_i}{\bar{D}}\hat{\tau}^{CBPS}\right\}^2$.\\

\subsection{Convergence rate under local misspecification}

\hspace{1em}Although the CBPS DID estimator is shown to have desirable properties in scenarios where at least one of the two working models is correctly specified, situations might arise where both of the working models are misspecified. To address this issue, \textcite{fan2022optimal} conduct a theoretical investigation of the AIPW ATE and CBPS ATE estimators under the scenario that both propensity score and outcome models are locally misspecified and find that the CBPS ATE estimator may converge in probability to the ATE at a faster rate than the AIPW estimator. In this subsection, we examine whether the CBPS DID estimator inherits such a desirable property in the DID design. We consider the same locally misspecified models as \textcite{fan2022optimal}:\\

\noindent\textbf{Assumption 4.}
 \begin{align*}
      &\pi(X_i)=\pi(X_i^\prime \beta^*)\exp(\xi u(X_i;\beta^*)), \tag{3.8} \label{locmis Assump beta} \\
      &m_{\Delta}(X_i)=X_i^\prime {\gamma^*} + \delta r(X_i), \tag{3.9} \label{locmis Assump gamma}
 \end{align*}
where $u(X_i;\beta^*)$ and $r(X_i)$ are functions determining the directions of misspecification with $| u(X_i;\beta^*)| \leq C$, $|r(X_i)| \leq C$ $a.s.$ for some positive constant $C$, $\xi \in \mathbb{R}$ and $\delta \in \mathbb{R}$ represent the magnitudes of misspecifications with $\xi=o(1)$ and $\delta=o(1)$.\\%, and $\beta^*$ and $\gamma^*$ is the approximate true value of $\beta$ and $\gamma$, respectively.\\
 
 \noindent\textbf{Theorem 4.} Under Assumptions 1-4 and Assumptions A, suppose at least one entry of \\
 $E\left[\left\{\frac{u(X_i;\beta^*)}{1-\pi(X_i^{\prime}{\beta}^*)}-\frac{\dot{\pi}(X_i^\prime \beta^*)X_i^{\prime}E\left[\frac{\dot{\pi}(X_i^\prime \beta^*)}{1-\pi(X_i^\prime \beta^*)}X_iX_i^{\prime}\right]^{-1}E\left[\frac{\pi(X_i^\prime \beta^*)}{1-\pi(X_i^\prime \beta^*)}u(X_i;\beta^*)X_i\right]}{1-\pi(X_i^{\prime}{\beta}^{*})}\right\}X_i\right]$ is nonzero, as $n\to \infty$, the CBPS DID and AIPW DID estimators satisfy:
\begin{equation}
     \hat{\tau}^{CBPS}-\tau=\frac{1}{n}\sum_{i=1}^{n}\eta_i^e+O_p({\xi^2\delta} +\delta n^{-1/2}+\xi n^{-1/2}), \tag{3.10} \label{CBPS asy rep locmis}
 \end{equation}
and
\begin{equation}
     \hat{\tau}^{AIPW}-\tau=\frac{1}{n}\sum_{i=1}^{n}\eta_i^e+O_p(\xi\delta +\delta n^{-1/2}+\xi n^{-1/2}), \tag{3.11} \label{AIPW asy rep locmis} 
\end{equation}
where $\eta_i^e$ is the efficient influence introduced in Theorem 1.\\

If $\sqrt{n}\xi\delta \to \infty$, then the CBPS DID estimator converges in probability to the ATT at a faster rate than the AIPW DID estimator since $\hat{\tau}^{CBPS}-\tau=O_p(n^{-1/2}+\xi^2\delta)$, whereas $\hat{\tau}^{AIPW}-\tau=O_p(\xi\delta)$. On the other hand, if $\sqrt{n}\xi\delta \to 0$, the two estimators have the same limiting distribution $N(0, E[\eta_i^{e2}])$, but $\sqrt{n}(\hat{\tau}^{CBPS}-\tau)$ converges to the limit distribution at a faster rate than $\sqrt{n}(\hat{\tau}^{AIPW}-\tau)$.
Theorem 4 implies that the CBPS DID estimator demonstrates greater robustness to slight model misspecification compared to the AIPW DID estimator. These faster rates of convergence are attributed to the arbitrariness of  $\gamma^{CBPS}$ in (\ref{CBPS Est Prop}), which effectively eliminates the product $\xi\delta$ in the asymptotic linear representation of the CBPS DID estimator. A detailed proof of this can be found in the Appendix.\\

%In this section we have conducted a thorough investigation of our CBPS DID estimator and show that it is locally semiparametric efficient when both working models are correctly specified and doubly robust in terms of consistency when one of two working model is misspecified, akin to the AIPW DID estimator. However the CBPS DID estimator outperforms the AIPW DID estimator by inheriting double robustness for inference when one of two working model is misspecified and a faster convergence rate when both working models are locally misspecified.\\

\section{Simulation}
\hspace{1em}In this section, we conduct a series of simulation studies to examine the finite sample properties of the CBPS DID estimator. The simulation designs here closely follow those in \parencites{sant2020doubly}{fan2022optimal}. Throughout these simulations, we utilize a logistic working model for the propensity score and a linear regression model for outcome evolution. %All observed covariates are incorporated linearly into these working models. 
For the OR, IPW, and AIPW approaches, we estimate the propensity scores using maximum likelihood estimation and estimate outcome evolution using ordinary least squares.

We set the sample size $n$ equal to 1000, and conduct 1000 Monte Carlo simulations for each design. The performance of various DID estimators is compared in terms of average bias, median bias, root mean square error (RMSE), empirical 95\% coverage probability, the average length of a 95\% confidence interval, and the average of their plug-in estimator for the asymptotic variance. The confidence intervals are constructed using the normal approximation, and the asymptotic variances are estimated by their sample analogues. Additionally, we present the semiparametric efficiency bound for each design calculated by Sant'Anna and Zhao (2020). This helps to assess the potential loss of efficiency or accuracy of a semiparametric DID estimator.\\

\subsection{Data generating process}
\hspace{1em} We conduct Monte Carlo simulations across five distinct scenarios as follow:\\
1. Both propensity score and outcome regression models are correctly specified.\\
2. Only the outcome regression model is correctly specified.\\
3. Only the propensity score model is correctly specified. \\
4. Both the propensity score and the outcome regression models are misspecified.\\
5. Both the propensity score and the outcome regression models are locally misspecified.\\

For a generic $W_i=(W_{1i},W_{2i},W_{3i},W_{4i})^{\prime}$, let
\begin{align*}
    &f_{or}(W_i)=210+27.4W_{1i}+13.7(W_{2i}+W_{3i}+W_{4i}),\\
    &f_{ps}(W_i)=0.75(-W_{1i}+0.5W_{2i}-0.25W_{3i}-0.1W_{4i}).
\end{align*}
Let $X_i=(X_{1i},X_{2i},X_{3i},X_{4i})^{\prime}$ be generated from $N(0,I_4)$, and $I_4$ be the $4 \times 4$ identity matrix. For $j=1,2,3,4$, let $Z_{ji}=(\Tilde{Z}_{ji}-E[\Tilde{Z}_{ji}])/\sqrt{Var(\Tilde{Z}_{ji})}$, where ${\Tilde{Z}_{1i}}=\exp(0.5X_{1i}), \Tilde{Z}_{2i}=10+X_{2i}/(1+\exp(X_{1i})),\Tilde{Z}_{3i}=(0.6+X_{1i}X_{3i}/25)^3$ and $\Tilde{Z}_{4i}=(20+X_{2i}+X_{4i})^2$. We consider the following data generating processes (DGPs):\\

DGP1.(PS and OR are correctly specified)
   \begin{align*}
&Y_{i1}(d)=2f_{or}(Z_i)+v(Z_i,D_i)+\varepsilon_{i1}(d),& &Y_{i0}(0)=f_{or}(Z_i)+v(Z_i,D_i)+\varepsilon_{i0},&\\
&\pi(Z_i)=\exp(f_{ps}(Z_i))/(1+\exp(f_{ps}(Z_i))),& &D_i=1\{\pi(Z_i)\geq U_i\},&\\
\end{align*}

DGP2.(PS is misspecified but OR is correctly specified)
    \begin{align*}
&Y_{i1}(d)=2f_{or}(Z_i)+v(Z_i,D_i)+\varepsilon_{i1}(d),& &Y_{i0}(0)=f_{or}(Z_i)+v(Z_i,D_i)+\varepsilon_{i0},& \\
&\pi(X_i)=\exp(f_{ps}(X_i))/(1+\exp(f_{ps}(X_i))), & &D_i=1\{\pi(X_i)\geq U_i\},&\\
\end{align*}

DGP3.(PS is correctly specified but OR is misspecified)
   \begin{align*}
&Y_{i1}(d)=2f_{or}(X_i)+v(X_i,D_i)+\varepsilon_{i1}(d),& &Y_{i0}(0)=f_{or}(X_i)+v(X_i,D_i)+\varepsilon_{i0},&\\
&\pi(Z_i)=\exp(f_{ps}(Z_i))/(1+\exp(f_{ps}(Z_i))), & &D_i=1\{\pi(Z_i)\geq U_i\},&\\
\end{align*}

DGP4.(PS and OR are misspecified)
   \begin{align*}
&Y_{i1}(d)=2f_{or}(X_i)+v(X_i,D_i)+\varepsilon_{i1}(d),&
&Y_{i0}(0)=f_{or}(X_i)+v(X_i,D_i)+\varepsilon_{i0},&\\
&\pi(X_i)=\exp(f_{ps}(X_i))/(1+\exp(f_{ps}(X_i))), & &D_i=1\{\pi(X_i)\geq U_i\},&\\
\end{align*}

DGP5.(PS and OR are locally misspecified)
\begin{align*}
&Y_{i1}(d)=2f_{or}(Z_i)+v(Z_i,D_i)+\varepsilon_{i1}(d)+2\delta r(Z_i),& &Y_{i0}(0)=f_{or}(Z_i)+v(Z_i,D_i)+\varepsilon_{i0}+\delta r(Z_i),&\\
&\pi(Z_i)=\exp(f_{ps}(Z_i))/(1+\exp(f_{ps}(Z_i))) \cdot \exp(\xi u(Z_i)),& &D_i=1\{\pi(Z_i)\geq U_i\},&\\
\end{align*}
where $d=0,1$ is an indicator of potential outcome, $D_i=0,1$ denotes whether unit $i$ receives treatment or not, $\varepsilon_{i0}$ and $\varepsilon_{i1}(d)$ are independent standard normal random variable, $U_i$ is an independent standard uniform random variable. For a generic $W_i$, $v(W_i,D_i)$ represents an independent normal random variable with a mean of $D_i \cdot f_{or}(W_i)$ and a variance of one. This $v$ serves as a proxy for time-invariant unobserved heterogeneity. $\xi=\delta=n^{-0.5}$ denotes the magnitudes of misspecification, $u(Z_i)=-Z_{1i}^2+Z_{2i}^2$ and $r(Z_i)=2Z_{1i}^2+4Z_{2i}^2+3Z_{3i}^2+Z_{4i}^2$ determine the directions of misspecification. It is important to note that in all the DGPs mentioned above,, the true ATT is zero, and the available data are $\{Y_{i0},Y_{i1},D_i,Z_i\}_{i=1}^n$, where $Z_i=(1,Z_{1i},Z_{2i},Z_{3i},Z_{4i})^{\prime}$ includes a constant among the covariates, the realized outcome$Y_{i0}$ and $Y_{i1}$ are generated according to $Y_{i0}=Y_{i0}(0)$ and $Y_{i1}=D_iY_{i1}(1)+(1-D_i)Y_{i1}(0)$ respectively.
 
\subsection{Results}
\hspace{1em} The results are summarized in the tables below, $\hat{\tau}^{IPW}$ represents the IPW DID estimator (\ref{IPW est}), $\hat{\tau}^{OR}$ denotes the OR DID estimator (\ref{OR est}), $\hat{\tau}^{AIPW}$ is the AIPW DID estimator (\ref{AIPW est}), and $\hat{\tau}^{CBPS}$ refers to the CBPS DID estimator (\ref{CBPS est}). The abbreviations used are as follows: ``Av.Bias'', ``Med.Bias'', ``RMSE'', ``Asy.V'', ``Cover'' and ``CIL'' stand for the average simulated bias, median simulated bias, simulated root mean-squared error, average of the plug-in estimators for the asymptotic variance, 95\% coverage probability, and the average length of the 95\% confidence interval. \\

\begin{table}[ht!]
\centering
%\captionsetup{justification=raggedright,singlelinecheck=false}
\caption{DGP1, both working models are correctly specified}
\begin{tabular}{l c c c c c c } 
 \hline
 \multicolumn{4}{c}{Semiparametric efficiency bound:11.1}\\
 \cline{2-7}
                      &Av.Bias  &Med.Bias &RMSE    &Asy.V   &Cover  &CIL       \\ [0.5ex]   
 \hline \\
$\hat{\tau}^{IPW}$    &0.091    &0.217    &2.805   &8403.740  &0.946  &10.526   \\
$\hat{\tau}^{OR}$     &0.002    &0.002    &0.101   &10.244    &0.954  &0.396     \\
$\hat{\tau}^{AIPW}$   &0.002    &0.003    &0.105   &11.245    &0.946  &0.414    \\
$\hat{\tau}^{CBPS}$   &0.002    &0.003    &0.105   &10.945    &0.943  &0.409    \\ [1ex] 
  \hline
\end{tabular}
\label{table 1}
\end{table}
Table 1 suggests that when both working models are correctly specified, all semiparametric DID estimators show minimal Monte Carlo bias. However, $\hat{\tau}^{OR}$, $\hat{\tau}^{AIPW}$, and $\hat{\tau}^{CBPS}$ outperform $\hat{\tau}^{IPW}$ in terms of bias, root mean square error, asymptotic variance, and the length of the confidence interval. This implies that the IPW DID estimator is substantially less efficient compared to the latter three. Although $\hat{\tau}^{OR}$ tends to be slightly more efficient than $\hat{\tau}^{AIPW}$ and $\hat{\tau}^{CBPS}$, the performance of these three estimators is quite similar.

\begin{table}[ht!]
\centering
%\captionsetup{justification=raggedright,singlelinecheck=false}
\caption{DGP2, correct outcome regression model with a misspecified propensity score model}
\begin{tabular}{l c c c c c c } 
 \hline
 \multicolumn{4}{c}{Semiparametric efficiency bound:11.6}\\
 \cline{2-7}
                      &Av.Bias  &Med.Bias &RMSE     &Asy.V     &Cover   &CIL       \\ [0.5ex]   
 \hline \\
$\hat{\tau}^{IPW}$    &2.068    &2.099    &3.297    &7472.489    &0.833   &10.072   \\
$\hat{\tau}^{OR}$     &0.003    &0.002    &0.100    &10.139      &0.947   &0.394     \\
$\hat{\tau}^{AIPW}$   &0.003    &0.004    &0.102    &10.459      &0.944   &0.400    \\
$\hat{\tau}^{CBPS}$   &0.003    &0.004    &0.103    &10.713      &0.947   &0.405    \\ [1ex] 
  \hline
\end{tabular}
\label{table 2}
\end{table}
Table 2 illustrates that when the propensity score model is misspecified, the CBPS DID estimator, $\hat{\tau}^{CBPS}$, is competitive with $\hat{\tau}^{OR}$ and $\hat{\tau}^{AIPW}$. As anticipated, $\hat{\tau}^{IPW}$ is biased in this scenario. Conversely, Table 3 indicates that when the outcome regression model is misspecified, $\hat{\tau}^{CBPS}$ outperforms the other three estimators in terms of root mean square error, asymptotic variance, and coverage probability. In this scenario, $\hat{\tau}^{OR}$ displays a non-negligible bias, which aligns with expectations.

\begin{table}[ht!]
\centering
%\captionsetup{justification=raggedright,singlelinecheck=false}
\caption{DGP3, misspecified outcome regression model with a correct propensity score model}
\begin{tabular}{l c c c c c c } 
 \hline
 \multicolumn{4}{c}{Semiparametric efficiency bound:11.1}\\
 \cline{2-7}
                      &Av.Bias   &Med.Bias  &RMSE     &Asy.V    &Cover   &CIL       \\ [0.5ex]   
 \hline \\
$\hat{\tau}^{IPW}$    &0.121     &0.313     &3.173    &10454.468  &0.941  &11.910   \\
$\hat{\tau}^{OR}$     &-1.352    &-1.330    &1.822    &1506.947   &0.826  &4.804     \\
$\hat{\tau}^{AIPW}$   &-0.001    &0.010     &1.223    &1834.486   &0.966  &5.234    \\
$\hat{\tau}^{CBPS}$   &-0.022    &0.002     &1.011    &977.372    &0.947  &3.870    \\ [1ex] 
  \hline
\end{tabular}
\label{table 3}
\end{table}

\begin{table}[ht!]
\centering
%\captionsetup{justification=raggedright,singlelinecheck=false}
\caption{DGP4, both models are misspecified}
\begin{tabular}{l c c c c c c } 
 \hline
 \multicolumn{4}{c}{Semiparametric efficiency bound:11.6}\\
 \cline{2-7}
                      &Av.Bias   &Med.Bias  &RMSE     &Asy.V    &Cover  &CIL       \\ [0.5ex]   
 \hline \\
$\hat{\tau}^{IPW}$    &-1.046    &-1.013    &2.609    &6092.271   &0.954  &9.243   \\
$\hat{\tau}^{OR}$     &-5.224    &-5.195    &5.372    &1472.513   &0.006  &4.751     \\
$\hat{\tau}^{AIPW}$   &-3.242    &-3.220    &3.494    &2674.697   &0.378  &6.240    \\
$\hat{\tau}^{CBPS}$   &-2.547    &-2.528    &2.727    &974.912    &0.265  &3.865    \\ [1ex] 
  \hline
\end{tabular}
\label{table 4}
\end{table}
In Table 4, when both working models are misspecified, it is unsurprising that all semiparametric DID estimators exhibit bias, and generally, the inference procedures are misleading. In this scenario, the CBPS DID estimator demonstrates smaller biases, lower root mean square error (RMSE), reduced asymptotic variance, and shorter confidence interval lengths compared to the OR and AIPW DID estimators. However, in DGP4, the IPW DID estimator appears to perform the best.

\begin{table}[ht!]
\centering
%\captionsetup{justification=raggedright,singlelinecheck=false}
\caption{DGP5, both models are locally misspecified}
\begin{tabular}{l c c c c c c } 
 \hline
 \multicolumn{4}{c}{Semiparametric efficiency bound:11.1}\\
\cline{2-7}
                      &Av.Bias  &Med.Bias  &RMSE     &Asy.V    &Cover  &CIL       \\ [0.5ex]   
 \hline \\
$\hat{\tau}^{IPW}$    &7.612    &7.477     &8.036    &6938.706   &0.118  &10.101   \\
$\hat{\tau}^{OR}$     &0.215    &0.209     &0.244    &12.587     &0.533  &0.439    \\
$\hat{\tau}^{AIPW}$   &0.118    &0.113     &0.166    &12.009     &0.777  &0.428    \\
$\hat{\tau}^{CBPS}$   &0.086    &0.083     &0.146    &13.032     &0.866  &0.446    \\ [1ex] 
  \hline
\end{tabular}
\label{table 5}
\end{table}
Table 5 indicates that in scenarios when both working models are locally misspecified, $\hat{\tau}^{CBPS}$ shows the smallest bias and root mean square error (RMSE), along with the best coverage probability. However, in DGP5, $\hat{\tau}^{IPW}$ demonstrates a non-negligible bias compared to the other three DID estimators. This finding corroborates the finding that IPW-based estimators are sensitive to even slight misspecifications of the propensity score model, \see \textcite{kang2007demystifying}.

In summary, the findings presented in Table 1 indicate that the estimated variance of $\hat{\tau}^{CBPS}$ is very close to the semiparametric efficiency bound when both working models are correctly specified. This supports our Theorem 1 regarding local semiparametric efficiency. In Tables 2 and 3, when one of the working models is misspecified, our proposed CBPS DID estimator  shows little bias, justifies the double robustness in terms of consistency as written in Theorem 2. Furthermore, in DGP2 and DGP3, the CBPS DID estimator achieves a coverage probability closer to 95\% compared to the AIPW DID estimator, validating the superiority of double robustness for inference demonstrated in Theorem 3. Lastly, Table 5 reveals that the CBPS DID estimator is more robust to mild model misspecification than the AIPW DID estimator, confirming the results of Theorem 4.\\

\section{An empirical application}
\hspace{1em} In this section, we apply our CBPS DID estimator to a real data sample. 
\textcite{lalonde1986evaluating} conducted a highly influential study evaluating the performance of different treatment effect estimators based on observational data. The study focused on whether a new statistical methodology could replicate an experimental benchmark: the treatment effect of a National Supported Work (NSW) labor training program on post-treatment earnings. Unfortunately, the results were not satisfactory due to the potential presence of selection bias in the observational data. Later, \textcite{dehejia1999causal}cdemonstrated that propensity score matching (PSM) based estimators could closely replicate the experimental results. However, \textcite{smith2005does} found that cross-sectional PSM estimators are highly sensitive to both model misspecification and sample selection. They suggested that DID matching estimators were more appropriate.

Following the findings of \textcite{smith2005does}, we use different samples and specifications to evaluate the existing DID estimators. Specifically, we utilize data from the Current Population Survey (CPS) to create a comparison group and use the control group from LaLonde's original experimental sample and the Dehejia and Wahba (DW) sample as our pseudo treatment group. We consider two datasets: (1) LaLonde's control group (425) + CPS (15,992), and (2) DW control group (260) + CPS (15,992). Since no one received training under this setup, the true ATT, if consistent, should be zero. Therefore, we use deviations from zero to evaluate the performance of the DID estimators.

The pre-treatment covariates in the data include age, real earnings in 1974, years of education, and dummy variables for high school dropout status, marital status, race (black), and ethnicity (Hispanic). The outcome of interest is real earnings in 1978, and pre-treatment real earnings in 1975 are also available. As part of our analysis, similar to Monte Carlo simulations, we compare the performance of the CBPS DID estimator,
$\hat{\tau}^{CBPS}$ with the IPW DID estimator, $\hat{\tau}^{IPW}$, the OR DID estimator, $\hat{\tau}^{OR}$, and the AIPW DID estimator, $\hat{\tau}^{AIPW}$. We assume that the outcome models are linear in parameters and the propensity score model follows a logistic specification.
To assess the sensitivity to model misspecification, we also consider three different specifications: (i) linear covariates (Lin); (ii) addition of some quadratic covariates such as age squared and education squared (Qua); (iii) addition of some interaction terms selected by SantAnna and Zhao (S\&Z). The results are summarized in Table 6, with standard error reported in parentheses. 

 \begin{table}[ht!]
 \centering
%\captionsetup{justification=raggedright,singlelinecheck=false}
\caption{Deviation of different DID estimators for the effect of training on real earnings in 1978, with CPS comparison group}
\begin{tabular}{l c c c c c c c c} 
 \hline
 \multicolumn{5}{c}{Lalonde Sample} & \multicolumn{4}{c}{DW Sample}\\
 \multicolumn{5}{c}{True ATT=0} & \multicolumn{4}{c}{True ATT=0}\\
 \hline
 \\
                      &$\hat{\tau}^{IPW}$  &$\hat{\tau}^{OR}$  &$\hat{\tau}^{AIPW}$     &$\hat{\tau}^{CBPS}$   &$\hat{\tau}^{IPW}$  &$\hat{\tau}^{OR}$  &$\hat{\tau}^{AIPW}$     &$\hat{\tau}^{CBPS}$\\ [0.5ex]   
 \hline \\
    Lin.              &-1310  &-1469  &-972    &-1030    &-397   &-560    &69      &2\\
                      &(424)  &(348)  &(415)   &(398)    &(593)  &(413)   &(566)   &(519)\\
                      \\
    Qua.              &-878   &-1248  &-746    &-744     &-407   &-277    &151     &211\\
                      &(551)  &(354)  &(525)   &(477)    &(1018) &(421)   &(884)   &(663)\\
                      \\
    S\&Z.             &-778   &-1426  &-717    &-730     &-229   &-676    &313     &239\\
                      &(523)  &(352)  &(507)   &(474)    &(937)  &(426)   &(825)   &(648)\\[1ex]
  \hline
\end{tabular}
\label{table 6}
\end{table}

The results in Table 6 reveal several interesting findings. First, $\hat{\tau}^{OR}$ displays the largest bias across different datasets and  covariates specification. For the Lalonde sample, every DID estimator shows more severe bias compared to their performance under the DW sample. Second, Abadie's IPW DID estimator $\hat{\tau}^{IPW}$ tends to have the largest standard error in all situations although its bias is relatively small especially under the Qua and S\&Z specifications. Third, $\hat{\tau}^{AIPW}$ and $\hat{\tau}^{CBPS}$ perform better than the other two in terms of bias. The CBPS DID estimator $\hat{\tau}^{CBPS}$ is very close to the true ATT when adopting the linear specification under the DW sample.  Finally, when we compare $\hat{\tau}^{CBPS}$ with $\hat{\tau}^{AIPW}$ we find that the CBPS DID estimator tends to have smaller standard errors in all situations. Taken together, the results suggest that the proposed DID estimator is a compelling alternative to existing DID estimators. Additionally, we use the Panel Study of Income Dynamics (PSID) to create a comparison group, with the results provided in the Appendix.

\section{Conclusion}
\hspace{1em}In this paper, we introduced an ATT estimator based on the CBPS method within a DID framework. This framework is applicable when the parallel trends assumption holds after conditioning on a set of pre-treatment covariates and when panel data are available. We conducted a thorough theoretical investigation of the CBPS DID estimator. Specifically, we found that while the CBPS DID estimator's expression is similar to that of the IPW estimator, its theoretical properties align more closely with those of the AIPW estimator. We demonstrated that the CBPS DID estimator is locally semiparametrically efficient and exhibits double robustness, similar to the AIPW DID estimator. Moreover, the asymptotic linear representation of the CBPS DID estimator remains invariant even when one of the working models is misspecified, a feature we refer to as double robustness for inference. Additionally, our estimator has a faster convergence rate than the AIPW DID estimator when both working models are locally misspecified. These superior properties set the CBPS DID estimator apart from the AIPW DID estimator. Our simulation results and empirical studies confirm these theoretical properties, showcasing the advantages of the proposed CBPS DID estimator.

In this work, we have primarily concentrated on the theoretical development of the CBPS DID estimator, especially in comparison to the AIPW DID estimator. An intriguing extension involves adapting the CBPS DID estimator to high-dimensional settings. This is a nontrivial task, as traditional regression methods tend to break down in high-dimensional settings, and CBPS methodology is no exception. To overcome this challenge, recent research has incorporated machine learning techniques into the first-step estimation of the propensity score and the outcome evolution. This approach, known as double machine learning methodology, has been explored in studies including \parencites{chernozhukov2017double}{chernozhukov2018double}{chang2020double}. Our ongoing research aims to develop and investigate a high-dimensional CBPS DID estimator.

\printbibliography

@Control{biblatex-control,
  options = {3.10:0:0:1:0:1:1:0:0:1:0:2:3:1:999:1:0:0:3:1:79:+:+:nyt},
}
	\clearpage

\appendix
%dummy comment inserted by tex2lyx to ensure that this paragraph is not empty%dummy comment inserted by tex2lyx to ensure that this paragraph is not empty

\section{Mathematical appendix}

\small

\section*{A1. Additional assumptions for first step estimations}\label{App1}
For simplify the notation, let $g(x)$ be a generic notation for $\pi(x)$ and $m_{\Delta}(x)$, denote a parametric function $g(x^\prime \theta)$ that serves as a generic representation for $\pi(x^\prime \beta)$ and $x^\prime \gamma$. Additionally, for a generic $W$, let $\|W\|=\sqrt{\text{trace}\left(W^\prime W\right)}$ denote the Euclidean norm of $W$.

Let 
\begin{equation*}
   \lambda_i(\zeta)= \frac{D_i-\pi(X_i^\prime \beta)}{E[D_i](1-\pi(X_i))}(\Delta Y_i - X_i^\prime \gamma),\ \  \Dot{\lambda}_i(\zeta) =\frac{\partial \lambda_i(\zeta)}{ \partial \zeta},\\
\end{equation*}
where $\zeta=(\beta^\prime, \gamma^\prime)^\prime$.\\

\noindent \textbf{Assumption A.} \\

\noindent(i) Assume there is a known parametric function $g(x^\prime \theta)=g(x)$ and $\theta \in \Theta \subset \mathbb{R}^K$, where $\Theta$ is a compact parameter set.\\

\noindent(ii) Assume $g(X_i^\prime \theta)$ is $a.s.$ continuous in $\theta \in \Theta$ and is $a.s.$ continuously differentiable, in addition, assume $\Dot{\pi}(X_i^\prime \beta)$ is bounded away from zero and infinity.\\ 

\noindent(iii) There exists a unique pseudo-true parameter $\theta^0 \in int\left(\Theta\right)$.\\ 

%\noindent(iv) Assume $E[X_iX_i^\prime|D_i=0]$ (used in OLS), $E\left[\frac{\left(D_i-2\pi(X_i^ \prime \beta^0)D_i +\pi^2(X_i^ \prime \beta^0)\right)\Dot{\pi}^2 (X_i^\prime \beta^0)}{\pi^2(X_i^ \prime \beta^0)(1-\pi(X_i^ \prime \beta^0)^2)}X_i X_i^\prime\right]$ (used in MLE) and $E\left[\frac{(1-D_i)\Dot{\pi}(X_i^\prime \beta^0)}{(1-\pi(X_i^\prime \beta^0))^2} X_i X_i^\prime\right]$ (used in the gamma and beta of CBPS) are non-singular.\\

\noindent(iv) Assume that $\hat{\theta}$ strongly converges to $\theta^0$ and can be asymptotically expressed as: 
\begin{equation*}
    \sqrt{n}(\hat{\theta}-\theta^0)=\frac{1}{\sqrt{n}}\sum_{i=1}^{n}\psi_{g}(D_i,\Delta Y_i, X_i,\theta^0)+o_p(1),
\end{equation*}
where $E\left[\psi_{g}(D_i,\Delta Y_i, X_i,\theta^0)\right]=0$, $E\left[\psi_{g}(D_i,\Delta Y_i, X_i,\theta^0) \psi_{g}(D_i,\Delta Y_i, X_i,\theta^0)^\prime\right]$ is finite and is positive definite.\\

\noindent(v) Assume that $E\left[\|\lambda_i(\zeta^0)\|^2\right] \textless \infty$ and $E\left[\text{sup}_{\zeta \in \Gamma^0} | \Dot{\lambda}_i(\zeta) |\right] \textless \infty$, where $\Gamma^0$ is a small neighborhood of $\zeta^0$.\\

Assumption A  (i)-(iv) represent standard conditions necessary for consistency and $\sqrt{n}$-asymptotically linear representations of the first-step estimators, and (v) imposes some weak integrability conditions. See Sant'Anna and Zhao (2020) for more details.

\section*{A2. ATT in OR approach}\label{App2}
\begin{align*}
&\tau=E\left[Y_{i1}(1)-Y_{i1}(0)|D_i=1\right]\\
&=E\left[Y_{i1}(1)-Y_{i0}(0)|D_i=1\right]-E\left[E\left[Y_{i1}(0)-{Y_{i0}(0)}|X_i,D_i=1\right]|D_i=1\right]\\
&=E\left[Y_{i1}(1)-Y_{i0}(0)|D_i=1\right]-E\left[E\left[Y_{i1}(0)-{Y_{i0}(0)}|X_i,D_i=0\right]|D_i=1\right] \ (\mbox{Assumption 2})\\
&=E\left[\Delta Y_i|D_i=1\right]-E\left[E\left[\Delta Y_i|X_i,D_i=0\right]|D_i=1\right]\\
&=E\left[\Delta Y_i|D_i=1\right]-E\left[m_{\Delta}(X_i)|D_i=1\right]=\tau^{OR}.
\end{align*}

\section*{A3. ATT in IPW approach}\label{App3}
\begin{align*}
    \tau^{IPW}&=E\left[\frac{D_i-\pi(X_i)}{E[D_i](1-\pi(X_i))}\Delta Y_i\right]\\
    &=\frac{E\left[D_i\Delta Y_i\right]}{E[D_i]}-\frac{E\left[\frac{(1-D_i)\pi(X_i)}{1-\pi(X_i)}\Delta Y_i\right]}{E[D_i]}\\
    &=\frac{E\left[D_i\Delta Y_i\right]}{\Pr(D_i=1)}-\frac{E\left[\frac{\pi(X_i)}{1-\pi(X_i)}E\left[(1-D_i)\Delta Y_i|X_i\right]\right]}{\Pr(D_i=1)}\\
    &=E\left[\Delta Y_i|D_i=1\right]-\frac{E\left[\frac{\pi(X_i)}{1-\pi(X_i)}E\left[1-D_i|X_i\right]E\left[Y_{i1}(0)-Y_{i0}(0)|X_i\right]\right]}{\Pr(D_i=1)} \ (\mbox{Assumption 2})\\
    &=E\left[\Delta Y_i|D_i=1\right]-\frac{E\left[\pi(X_i)E\left[\Delta Y_i|X_i,D_i=0\right]\right]}{\Pr(D_i=1)}\\
    &=E\left[\Delta Y_i|D_i=1\right]-E\left[E\left[\Delta Y_i|X_i,D_i=0\right]|D_i=1\right]=\tau
\end{align*}

\section*{A4. ATT in AIPW approach}\label{App4}
\begin{align*}
    \tau^{AIPW}&=E\left[\frac{D_i-\pi(X_i)}{E[D_i](1-\pi(X_i))}(\Delta Y_i-m_{\Delta}(X_i))\right]\\
    &=E\left[\frac{D_i-\pi(X_i)}{E[D_i](1-\pi(X_i))}\Delta Y_i\right]-E\left[\frac{E[D_i|X_i]-\pi(X_i)}{E[D_i](1-\pi(X_i))}m_{\Delta}(X_i)\right]\\
    &=\tau^{IPW}-0=\tau
\end{align*}

\section*{A5. Proof of Theorem 1}\label{App5}
Suppose that  both working models are correctly specified, that is $\pi(X_i^\prime \beta)=\pi(X_i)$, $X_i^\prime \gamma=m_{\Delta}(X_i)$. Then, by using the expression (\ref{CBPS Est Prop}), a Taylor expansion yields:
\begin{align*}
\sqrt{n}(\hat{\tau}^{CBPS}-\tau)
%&=\frac{1}{\sqrt{n}} \sum_{i=1}^{n}\left\{\frac{D_i-\pi(X_i^\prime \hat{\beta}^{CBPS})}{\bar{D}(1-\pi(X_i^\prime \hat{\beta}^{CBPS}))}(\Delta Y_i-X_i^\prime \gamma^{CBPS})-\frac{D_i}{\bar{D}}\tau \right\}\\
&=\frac{1}{\sqrt{n}} \sum_{i=1}^{n}\left\{\frac{D_i-\pi(X_i^\prime \beta)}{E[D_i](1-\pi(X_i^\prime \beta))}(\Delta Y_i - X_i^\prime \gamma^{CBPS})-\frac{D_i}{E[D_i]}\tau \right\}\\
&-\sqrt{n}(\bar{D}-E[D_i])E\left[\frac{(D_i-\pi(X_i^\prime \beta))(\Delta Y_i-X_i^\prime \gamma^{CBPS})}{E^2[D_i](1-\pi(X_i^\prime \beta))}-\frac{D_i}{E^2[D_i]}\tau\right]\\
&-\sqrt{n}(\hat{\beta}^{CBPS}-\beta)^\prime E\left[\frac{(1-D_i)(\Delta Y_i - X_i^\prime \gamma^{CBPS})\Dot{\pi}(X_i^\prime \beta)}{E[D_i](1-\pi(X_i^\prime \beta))^2}X_i\right]+o_p(1)\\
&=\frac{1}{\sqrt{n}} \sum_{i=1}^{n}\left\{\frac{D_i-\pi(X_i^\prime \beta)}{E[D_i](1-\pi(X_i^\prime \beta))}(\Delta Y_i - X_i^\prime \gamma^{CBPS})-\frac{D_i}{E[D_i]}\tau \right\}+o_p(1)\\
&=\frac{1}{\sqrt{n}} \sum_{i=1}^{n} \eta_i^e + o_p(1),
\end{align*}
where the second equality follows from 
\begin{align*}
E\left[\frac{(D_i-\pi(X_i^\prime \beta))(\Delta Y_i-X_i^\prime \gamma^{CBPS})}{E^2[D_i](1-\pi(X_i^\prime \beta))}-\frac{D_i}{E^2[D_i]}\tau\right]&= E\left[\frac{(D_i-\pi(X_i))(\Delta Y_i- m_{\Delta}(X_i))}{E^2[D_i](1-\pi(X_i))}-\frac{D_i}{E^2[D_i]}\tau\right]\\
&=\frac{\tau^{AIPW}-\tau}{E[D_i]}=0,\\
E\left[\frac{(1-D_i)(\Delta Y_i - X_i^\prime \gamma^{CBPS})\Dot{\pi}(X_i^\prime \beta)}{E[D_i](1-\pi(X_i^\prime \beta))^2}X_i\right]&=
     E\left[\frac{E\left[(1-D_i)(\Delta Y_i - m_{\Delta}(X_i))|X_i\right]\Dot{\pi}(X_i^\prime \beta)}{E[D_i](1-\pi(X_i))^2}X_i \right]=0
 \end{align*}
by setting $\gamma^{CBPS}=\gamma$. (Note that $\gamma^{CBPS}$ can take any value.) Thus the conclusion follows by CLT.
    
\section*{A6. Proof of Theorem 2}\label{App6}
We separate the proof into two cases.\\

\noindent Case 1: If the outcome model is correctly specified, that is $X_i^\prime{\gamma_0}=m_{\Delta}(X_i)$ but $\pi(X_i^\prime \beta_0)\neq \pi(X_i)$, by the weak law of large numbers and the continuous mapping theorem, as $n \to \infty$,
\begin{align*}
       \hat{\tau}^{CBPS}&\xrightarrow{p}E\left[\frac{D_i-\pi(X_i^{\prime}\beta_0)}{E[D_i](1-\pi(X_i^{\prime}\beta_0))}(\Delta Y_i-m_{\Delta}(X_i))\right]\\
        &=E\left[\frac{D_i\Delta Y_i}{E[D_i]}-\frac{D_i m_{\Delta}(X_i)}{E[D_i]}\right]
        -E\left[\frac{(1-D_i)(\Delta Y_i- m_{\Delta}(X_i))\pi (X_i^\prime \beta_0)}{E[D_i](1-\pi (X_i^\prime \beta_0))}\right]\\
        &=E\left[\frac{D_i\Delta Y_i}{E[D_i]}-\frac{D_i m_{\Delta}(X_i)}{E[D_i]}\right]
        -E\left[\frac{E[(1-D_i)(\Delta Y_i- m_{\Delta}(X_i))|X_i]\pi (X_i^\prime \beta_0)}{E[D_i](1-\pi (X_i^\prime \beta_0))}\right]\\
        &=\tau^{OR}-0=\tau,
 \end{align*}
where the first equality follows from $\frac{D_i-\pi(X_i^{\prime}\beta_0)}{1-\pi(X_i^{\prime}\beta_0)}=D_i-\frac{(1-D_i)\pi(X_i^{\prime}\beta_0)}{1-\pi(X_i^{\prime}\beta_0)}$ and the third equality follows from $E\left[(1-D_i)(\Delta Y_i-m_{\Delta}(X_i))|X_i\right]=0$.\\

\noindent Case 2: If only the propensity score model is correctly specified, that is, $\pi(X_i^\prime \beta_0)=\pi(X_i)$ but $X_i^\prime{\gamma_0}\neq m_{\Delta}(X_i)$, by the weak law of large numbers and the continuous mapping theorem, as $n \to \infty$,
\begin{align*}
        \hat{\tau}^{CBPS}&\xrightarrow{p}E\left[\frac{D_i-\pi(X_i)}{E[D_i](1-\pi(X_i))}(\Delta Y_i-X_i^\prime \gamma^{CBPS})\right]\\
        &=E\left[\frac{D_i-\pi(X_i)}{E[D_i](1-\pi(X_i))}\Delta Y_i\right]-E\left[\frac{E\left[D_i|X_i\right]-\pi(X_i)}{E[D_i](1-\pi(X_i))}X_i^\prime \gamma^{CBPS} \right]\\
        &=\tau^{IPW}-0=\tau,
    \end{align*}
where the second equality follows from $E[D_i|X_i]=\pi(X_i)$.\\

\section*{A7. Proof of Theorem 3}\label{App7}
By Taylor expansions, we have 
\begin{align*}
&\sqrt{n}(\hat{\tau}^{CBPS}-\tau) \\
  &= \frac{1}{\sqrt{n}}\sum_{i=1}^n \eta_{i}^{CBPS}
  -\sqrt{n}(\bar{D}-E[D_i])E\left[\frac{(D_i-\pi(X_i^\prime \beta_0^{CBPS}))(\Delta Y_i-X_i^\prime \gamma^{CBPS})}{E^2[D_i](1-\pi(X_i^\prime \beta_0^{CBPS}))}-\frac{D_i}{E^2[D_i]}\tau\right]\\
&-\sqrt{n}(\hat{\beta}^{CBPS}-\beta_0^{CBPS})^{\prime}{E\left[\frac{(1-D_i)(\Delta Y_i-X_i^{\prime}{\gamma^{CBPS}})\Dot{\pi}(X_i^{\prime}\beta_0^{CBPS})}{E[D_i](1-\pi(X_i^{\prime}\beta_0^{CBPS}))^2} X_i\right]}+o_p(1),\tag{A.1} \label{expansion of CBPS}
        \end{align*}
whereas
\begin{align*}
  &\sqrt{n}(\hat{\tau}^{AIPW}-\tau)\\
   & = \frac{1}{\sqrt{n}}\sum_{i=1}^n \eta_{i}^{AIPW}-
   \sqrt{n}(\bar{D}-E[D_i])E\left[\frac{(D_i-\pi(X_i^\prime \beta_0^{AIPW}))(\Delta Y_i-X_i^\prime \gamma_0^{AIPW})}{E^2[D_i](1-\pi(X_i^\prime \beta_0^{AIPW}))}-\frac{D_i}{E^2[D_i]}\tau\right]\\
    &-\sqrt{n}(\hat{\beta}^{AIPW}-\beta_0^{AIPW})^{\prime}{E\left[\frac{(1-D_i)(\Delta Y_i-X_i^{\prime}{\gamma_0^{AIPW}})\Dot{\pi}(X_i^\prime \beta_0^{AIPW})}{E[D_i](1-\pi(X_i^{\prime}\beta_0^{AIPW}))^2}X_i\right]}\\
       &-\sqrt{n}(\hat{\gamma}^{AIPW}-\gamma_0^{AIPW})^{\prime}{E\left[\frac{D_i-\pi(X_i^{\prime}\beta_0^{AIPW})}{E[D_i](1-\pi(X_i^{\prime}\beta_0^{AIPW}))}X_i\right]}+o_p(1). \tag{A.2} \label{expansion of AIPW}
\end{align*}
As in the proof of Theorem 2, we separate the proof into two cases.\\

\noindent Case 1: If the outcome regression model is correct but the propensity score model is incorrect, the third terms of both AIPW and CBPS expansions are zero but the fourth term of the AIPW expansion is nonzero and of order $O_p(1)$ because $\pi(X_i)\neq \pi(X_i^{\prime}\beta_0^{AIPW})$. \\

\noindent Case 2: On the other hand, if the propensity score model is correct but the outcome regression model is incorrect, the fourth term of the AIPW expansion is zero but the third terms of the AIPW expansion is nonzero and of order $O_p(1)$. However, the arbitrariness of $\gamma^{CBPS}$ in the CBPS expansion offers a significant advantage. Specifically, the third term of CBPS expansion is zero even when the outcome regression model is incorrect by setting $\gamma^{CBPS}=E\left[\frac{(1-D_i)\Dot{\pi}(X_i^\prime \beta_0^{CBPS})}{(1-\pi(X_i^{\prime}\beta_0^{CBPS}))^2} X_iX_i^\prime\right]^{-1}E\left[\frac{(1-D_i)\Delta Y_i\Dot{\pi}(X_i^\prime \beta_0^{CBPS})}{(1-\pi(X_i^{\prime}\beta_0^{CBPS}))^2}X_i\right] :=\gamma_0^{CBPS}$.\\

\section*{A8. Proof of Theorem 4}\label{App8}

We provide a sketch of the proof because the detail is very similar to that of (3.10) and (3.11) of Fan et al. (2021).
Letting $\beta_0^{CBPS}$ denote the probability limit of $\hat{\beta}^{CBPS}$, we decompose
\begin{align*}
&\hat{\tau}^{CBPS}-\tau\\
&=\frac{1}{\bar{D}}\frac{1}{n}\sum_{i=1}^n \left[\frac{D_i-\pi(X_i^{\prime}\beta_0^{CBPS})}{1-\pi(X_i^{\prime}\beta_0^{CBPS})}\left(\Delta Y_i-X_i^{\prime}\gamma^{CBPS}\right)-D_i\tau\right]\\
&+\frac{1}{\bar{D}}\frac{1}{n}\sum_{i=1}^n \left[\frac{D_i-\pi(X_i^{\prime}\hat{\beta}^{CBPS})}{1-\pi(X_i^{\prime}\hat{\beta}^{CBPS})}-\frac{D_i-\pi(X_i^{\prime}\beta_0^{CBPS})}{1-\pi(X_i^{\prime}\beta_0^{CBPS})}\right]\left(\Delta Y_i-X_i^{\prime}\gamma^{CBPS}\right)\\
&:= A_1+A_2.
\end{align*}

First, we write $\gamma^{CBPS}=\gamma^*+\delta A$ for any value $A$ since $\gamma^{CBPS}$ is arbitrary. 
Then we have
\begin{align*}
\Delta Y_i-X_i^{\prime}\gamma^{CBPS}
&=\Delta Y_i-m_{\Delta}(X_i)+\left\{m_{\Delta}(X_i)-X_i^{\prime}\gamma^{CBPS}\right\}\\
&=\Delta Y_i-m_{\Delta}(X_i)+\delta\left\{r(X_i)-X_i^{\prime}A\right\}.\tag{A.3}  \label{OR misspecification}
\end{align*}
Also, by the same argument as the proof of (C.1) in Fan et al. (2021), we have 
\begin{equation*}
\beta_0^{CBPS}-\beta^{*}=\xi E\left[\frac{\dot{\pi}_i^{*}}{1-\pi^{*}_i}X_iX_i^{\prime}\right]^{-1}E\left[\frac{\pi^{*}_i}{1-\pi^{*}_i}u_i^{*}X_i^{\prime}\right]+O(\xi^2).\tag{A.4}  \label{PS misspecification}
\end{equation*}
where $u_i^{*}=u_i(X_i; \beta^{*})$, $\pi_i^{*}=\pi(X_i^{\prime}\beta^{*})$ and $\dot{\pi}_i^{*}=\dot{\pi}(X_i^{\prime}\beta^{*})$.
Hence by using a similar argument as the proof of Theorem 1 with (\ref{OR misspecification}) and (\ref{PS misspecification}), we have 
\begin{equation*}
A_2=O_p(\delta n^{-1/2}).
\end{equation*}
and 
\begin{align*}
A_1&= \frac{1}{n}\sum_{i=1}^n \eta_i^e+\frac{1}{E[D_i]}E\left[\frac{D_i-\pi(X_i^{\prime}\beta_0^{CBPS})}{1-\pi(X_i^{\prime}\beta_0^{CBPS})}\left(m_{\Delta}X_i-X_i^{\prime}\gamma^{CBPS}\right)\right]+O_p(\xi n^{-1/2}+\delta n^{-1/2})\\
&= \frac{1}{n}\sum_{i=1}^n \eta_i^e+\frac{1}{E[D_i]}E\left[\frac{D_i-\pi(X_i^{\prime}\beta_0^{CBPS})}{1-\pi(X_i^{\prime}\beta_0^{CBPS})}\delta\left\{r(X_i)-X_i^{\prime}A\right\}\right]+O_p(\xi n^{-1/2}+\delta n^{-1/2})\\
&= \frac{1}{n}\sum_{i=1}^n \eta_i^e+O_p(\xi^2\delta+\xi n^{-1/2}+\delta n^{-1/2}).
\end{align*}
To see the last equality, the second term is calculated as 
\begin{align*}
&\frac{1}{E[D_i]}E\left[\frac{D_i-\pi(X_i^{\prime}\beta_0^{CBPS})}{1-\pi(X_i^{\prime}\beta_0^{CBPS})}\delta\left\{r(X_i)-X_i^{\prime}A\right\}\right]\\
&=\frac{1}{E[D_i]}E\left[\left\{\frac{D_i-\pi(X_i^{\prime}{\beta}^*)}{1-\pi(X_i^{\prime}{\beta}^*)}-\frac{(1-D_i)\dot{\pi}_i^{*}X_i^{\prime}(\beta_0^{CBPS}-\beta^*)}{\left\{1-\pi(X_i^{\prime}{\beta}^{*})\right\}^2}\right\}\delta\left\{r(X_i)-X_i^{\prime}A\right\}\right]+O(\xi^2\delta)\\
&=\frac{1}{E[D_i]}E\left[\left\{\frac{\pi(X_i^{\prime}\beta^{*})\left(1+\xi u_i^{*}\right)-\pi(X_i^{\prime}{\beta}^*)}{1-\pi(X_i^{\prime}{\beta}^*)}-\frac{\left\{1-\pi(X_i^{\prime}\beta^{*})\left(1+\xi u_i^{*}\right)\right\}\dot{\pi}_i^{*}X_i^{\prime}(\beta_0^{CBPS}-\beta^*)}{\left\{1-\pi(X_i^{\prime}{\beta}^{*})\right\}^2}\right\}\delta\left\{r(X_i)-X_i^{\prime}A\right\}\right]\\
&+O(\xi^2\delta)\\
&=\frac{1}{E[D_i]}E\left[\left\{\frac{\xi u_i^{*}}{1-\pi(X_i^{\prime}{\beta}^*)}-\frac{\xi\dot{\pi}_i^{*}X_i^{\prime}E\left[\frac{\dot{\pi}_i^{*}}{1-\pi_i^{*}}X_iX_i^{\prime}\right]^{-1}E\left[\frac{\pi_i^{*}}{1-\pi_i^{*}}u_i^{*} X_i\right]}{1-\pi(X_i^{\prime}{\beta}^{*})}\right\}\delta\left\{r(X_i)-X_i^{\prime}A\right\}\right]+O(\xi^2\delta)\\
&=\frac{\xi\delta}{E[D_i]}E\left[\left\{\frac{u_i^{*}}{1-\pi(X_i^{\prime}{\beta}^*)}-\frac{\dot{\pi}_i^{*}X_i^{\prime}E\left[\frac{\dot{\pi}_i^{*}}{1-\pi_i^{*}}X_iX_i^{\prime}\right]^{-1}E\left[\frac{\pi_i^{*}}{1-\pi_i^{*}}u_i^{*} X_i\right]}{1-\pi(X_i^{\prime}{\beta}^{*})}\right\}\left\{r(X_i)-X_i^{\prime}A\right\}\right]+O(\xi^2\delta).
\end{align*}
Hence assuming that at least one entry of $E\left[\left\{\frac{u_i^{*}}{1-\pi(X_i^{\prime}{\beta}^*)}-\frac{\dot{\pi}_i^{*}X_i^{\prime}E\left[\frac{\dot{\pi}_i^{*}}{1-\pi_i^{*}}X_iX_i^{\prime}\right]^{-1}E\left[\frac{\pi_i^{*}}{1-\pi_i^{*}}u_i^{*} X_i\right]}{1-\pi(X_i^{\prime}{\beta}^{*})}\right\}X_i\right]$ is nonzero, there exists $A$ such that
\begin{equation*}
E\left[\left\{\frac{u_i^{*}}{1-\pi(X_i^{\prime}{\beta}^*)}-\frac{\dot{\pi}_i^{*}X_i^{\prime}E\left[\frac{\dot{\pi}_i^{*}}{1-\pi_i^{*}}X_iX_i^{\prime}\right]^{-1}E\left[\frac{\pi_i^{*}}{1-\pi_i^{*}}u_i^{*} X_i\right]}{1-\pi(X_i^{\prime}{\beta}^{*})}\right\}\left\{r(X_i)-X_i^{\prime}A\right\}\right]=0.
\end{equation*} 
This completes the proof of (\ref{CBPS asy rep locmis}). The proof of (\ref{AIPW asy rep locmis}) follows from the same argument except that $\hat{\gamma}^{AIPW}$ is not arbitrary unlike $\gamma^{CBPS}$.
\clearpage

\section*{A9. Additional Application Results}\label{App9}

\begin{table}[ht!]
 \centering
%\captionsetup{justification=raggedright,singlelinecheck=false}
\caption{Deviation of different DID estimators for the effect of training on real earnings in 1978, with PSID comparison group}
\begin{tabular}{l c c c c c c c c} 
 \hline
 \multicolumn{5}{c}{Lalonde Sample} & \multicolumn{4}{c}{DW Sample}\\
 \multicolumn{5}{c}{True ATT=0} & \multicolumn{4}{c}{True ATT=0}\\
 \hline
 \\
                      &$\hat{\tau}^{IPW}$  &$\hat{\tau}^{OR}$  &$\hat{\tau}^{AIPW}$     &$\hat{\tau}^{CBPS}$   &$\hat{\tau}^{IPW}$  &$\hat{\tau}^{OR}$  &$\hat{\tau}^{AIPW}$     &$\hat{\tau}^{CBPS}$\\ [0.5ex]   
 \hline \\
    Lin.              &-1038  &-1605  &1190     &661      &2238   &-664    &4062     &1932\\
                      &(821)  &(699)  &(1169)   &(505)    &(1196) &(898)   &(2298)   &(552)\\
                      \\
    Qua.              &812    &-1236  &1357     &930      &1785   &-308    &4673     &2331\\
                      &(825)  &(704)  &(1177)   &(569)    &(1268) &(902)   &(2480)   &(681)\\
                      \\
    S\&Z.             &811    &-947   &1039     &915      &1750    &209     &3934     &2215\\
                      &(826)  &(642)  &(1052)   &(564)    &(1310)  &(812)   &(2118)   &(715)\\[1ex]
  \hline
\end{tabular}
\label{table 7}
\end{table}
 
\end{document}